\documentclass{article}
\usepackage{slashed}

\def\beq{\begin{eqnarray}}
\def\eeq{\end{eqnarray}}
\def\bea{\begin{eqnarray*}}
\def\eea{\end{eqnarray*}}

\def\mh{m^{(1)}_{hol}}
\def\mmh{m^{(2)}_{hol}}
\def\mih{m^{(i)}_{hol}}

\def\singleandthirdspaced{\baselineskip=\normalbaselineskip\multiply
    \baselineskip by 130\divide\baselineskip by 100}

\newcommand{\newc}{\newcommand}
\newc{\qbar}{{\overline q}}
\newc{\Kahler}{K\"ahler }
\newc{\deltaGS}{\delta_{\rm GS}}

\begin{document}
\begin{titlepage}
\begin{flushright}
{\large hep-th/yymmnnn \\
SCIPP 11/03\\
}
\end{flushright}

\vskip 1.2cm

\begin{center}

{\LARGE\bf Supersymmetric QCD: Exact Results and Strong Coupling}

\vskip 1.4cm

{\large  Michael Dine, Lawrence Pack, Chang-Soon Park, Lorenzo Ubaldi and Weitao Wu}
\\
\vskip 0.4cm
{\it Santa Cruz Institute for Particle Physics and
\\ Department of Physics,
     Santa Cruz CA 95064  } \\
\vskip 0.4cm
{\large Guido Festuccia}
\vskip 0.1cm
{\it  Institute for Advanced Study \\ Princeton, New Jersey  08540}
\vskip 4pt

\vskip 1.5cm

\begin{abstract}
We revisit two longstanding puzzles in supersymmetric gauge theories.  The first concerns the question of the holomorphy
of the coupling, and related to this the possible definition of an exact (NSVZ) beta function.  The second
concerns instantons in pure gluodynamics, which appear to give sensible,
exact results for certain correlation functions, which nonetheless differ from those obtained using systematic weak
coupling expansions.  For the first question, we extend an earlier proposal of Arkani-Hamed and Murayama, showing that if their
regulated action is written suitably, the holomorphy of the couplings is manifest, and it is easy to determine the renormalization scheme
for which the NSVZ formula holds.  This scheme, however, is seen to be one of an infinite class of
schemes, each leading to an exact beta function; the NSVZ scheme, while simple, is not selected by any compelling
physical consideration.  For the second question, we explain why the instanton computation in the
pure supersymmetric gauge theory is not reliable, even at short distances.  The semiclassical expansion about the instanton is purely formal; if infrared divergences appear,
they spoil arguments based on holomorphy.  We demonstrate that infrared divergences do not occur in the perturbation expansion
 about the instanton, but explain that there is no reason to think this captures all contributions from the sector with unit topological
 charge.  That one expects additional contributions is illustrated by dilute gas corrections.
 These are infrared divergent, and so difficult to define, but if non-zero give
 order one, holomorphic, corrections
to the leading result.    Exploiting an earlier analysis of Davies et al, we demonstrate that in the theory compactified on a circle
of radius $\beta$, due to infrared effects, finite contributions indeed arise which are not visible in the formal $\beta \rightarrow \infty$ limit.
\end{abstract}

\end{center}

\vskip 1.0 cm

\end{titlepage}
\setcounter{footnote}{0} \setcounter{page}{2}
\setcounter{section}{0} \setcounter{subsection}{0}
\setcounter{subsubsection}{0}

\singleandthirdspaced

\section{Two Puzzles}
\label{paradox}

In a rich research program, which stimulated much of the work of the past 25 years on supersymmetric dynamics, Novikov, Shifman, Vainshtein and Zakharov (NSVZ) studied instanton
effects in supersymmetric gauge theories\cite{Novikov:1983ee,Novikov:1983uc,Shifman:1999mv,amatireview}.
NSVZ considered $SU(N)$ gauge theories with and without chiral matter (gluodynamics).   In the case of pure supersymmetric $SU(2)$ (``SUSY Gluodynamics"), NSVZ considered the Green's function:
\beq
\Delta(x) \equiv\langle W_\alpha^2(x_1,\theta_1) W_\beta^2(x_2,\theta_2) \rangle \vert_{\theta_1 = \theta_2 = 0} = \langle \left(\lambda(x_1)\lambda(x_1)\right) \left(\lambda(x_2)
\lambda(x_2) \right)\rangle,
\eeq
where trace over gauge indices is understood.
They noted, first, that as a correlation function of chiral fields, the lowest component is necessarily a constant.  This follows since, for the correlator of a chiral
field, $\Phi(y) = A(y) + \sqrt{2} \psi(y) + \theta^2 F(y)$ with $y^{\mu} = x^{\mu}+ i \theta \sigma^{\mu} \bar\theta$, a transformation by $\bar \epsilon \bar Q$ on the correlator $\langle A(0) \psi(x) \rangle$  yields
\beq
\slashed{\partial} \langle A(0) A(x) \rangle =0.
\eeq
At short distance, NSVZ argued that, since the theory is asymptotically free, the Green's function can be reliably computed, provided that it is well-defined.
Remarkably, they found that the leading result is infrared finite and, indeed, a constant.  For the generalization of $\Delta$ to $SU(N)$:
\beq
\Delta(x) &=& \langle (\lambda(x_1)\lambda (x_1)) \cdots (\lambda (x_N) \lambda (x_N)) \rangle \nonumber \\
&=& \frac{2^N}{(N-1)! (3N-1)} M_{PV}^{3N} e^{- {8 \pi^2 \over g^2}- i \theta}{1 \over g^{2N}}.
\label{deltaequation}
\eeq
Here $M_{PV}$ is the mass of a Pauli-Villars regulator (of which much more shortly).

NSVZ then contended that the leading instanton result was exact, as a consequence of the symmetries
of the theory.  More precisely, by considering the structure of the collective coordinate
measure and the constraints on the correlator (for fixed values of the collective coordinates) due to
supersymmetry, they asserted that all Feynman graph corrections to the leading
order result would cancel.

Finally, since unbroken supersymmetry requires that $\Delta$ is constant, cluster decomposition, they argued, requires that
\beq
\langle \left(\lambda(x) \lambda(x)\right) \left(\lambda(0) \lambda(0)\right) \rangle = \langle \lambda(0) \lambda(0) \rangle^2,
\eeq
and the single instanton computation gives a reliable calculation of the gaugino condensate.
This analysis generalizes to $SU(N)$, where one studies a correlation function involving $N$ factors of $\lambda \lambda$, as in eqn.(\ref{deltaequation}).

This result also
leads to the interesting proposal that one can write an {\it exact} beta function in supersymmetric gauge theories\cite{exactbeta}\footnote{This beta function was first
written down by Jones in an early study of the multiplet of anomalies\cite{jones}.}.
NSVZ argued that the instanton result, including the one loop determinant, was exact and renormalization group invariant.
Differentiating eqn.(\ref{deltaequation}) then yields:
\beq
\beta(g) = -{3N{g^3 \over 16 \pi^2} \over 1 - 2N{g^2 \over 16 \pi^2}}.
\eeq

These two results are closely related, as we can see by considering the consequences of holomorphy for $\Delta(x).$
Indeed, the NSVZ discussion is a precursor of Seiberg's exploitation of holomorphy\cite{Seiberg:1994bp}.
As first stressed by NSVZ, a chiral correlator, such as $\Delta$, if well-defined, is necessarily a holomorphic function of
\beq
\tau = {8\pi^2 \over g_{hol}^2} + i {\theta}.
\eeq
Care, as they noted, is required in the definition of $g_{hol}$ in this expression.  Redefinitions of $g$ change the real
part of $\tau$.  They argued that, indeed, the holomorphic coupling was related to a more conventional definition by
\beq
{8 \pi^2 \over g_{hol}^2} = {8 \pi^2 \over g^2} +2N \log(g).
\label{nsvzconnection}
\eeq
Understanding this connection will be much of the focus of this paper.
The single instanton amplitude is proportional to $e^{i \theta}$.
As the perturbative corrections to the instanton introduce no further $\theta$-dependence, holomorphy requires that the result must be proportional
to $e^{-\tau}$.\footnote{More precisely, since the theory
is invariant under $\theta \rightarrow \theta + 2\pi$, only the $\tau$-dependence $e^{-n \tau}$ is allowed in holomorphic quantities.}
The instanton amplitude, at one loop, in the regularization scheme introduced by 't Hooft, is given by
\beq
M_{PV}^{3N}{1 \over g^{2N}} e^{-{8 \pi^2 \over g^2} - i\theta}.
\eeq
This combination is precisely $\Lambda^{3N}$, where $\Lambda$ is the conventional renormalization group invariant
 scale, computed through two loop order.  The NSVZ assertion that the result is exact is the statement that $\Lambda$ is given
 exactly by this expression, coinciding with the $\beta$ function of NSVZ.

While carefully reasoned, each of these results raises interesting questions.  For the exact beta function, it is natural to ask:  what, precisely, is the scheme
in which this expression holds?  By considering an abelian gauge theory, Shifman and Vainshtein provided a conceptual setting\cite{svabelian}.  In this context one can
introduce a supersymmetric Pauli-Villars regulator.  The mass of the regulator field is a superpotential term; written in terms of this ``bare" mass parameter
the effective action should be holomorphic.  However, the kinetic term for the regulator field is renormalized, and the ``physical" mass scale is related to the
bare mass term by a wave function renormalization factor.  This permits the definiton of an ``exact" beta function, written in terms of an anomalous dimension which must
be computed order by order in perturbation theory.  By comparing theories with different regulator masses, $m_1$ and $m_2$, it is possible to define
a precise notion of a Wilsonian action, integrating out between $m_1$ and $m_2$\cite{dineshirman}.  This action is holomorphic in the bare, holomorphic masses, and
the Wilsonian evolution of couplings is exhausted at one loop.  In terms of physical
mass scales, the couplings evolve with more conventional renormalization group equations.

The non-abelian case is complicated, in general, by the problem of writing a holomorphic regulator.
The answer often given as to scheme dependence is that the regularization/renormalization scheme leading to the
NSVZ $\beta$-function is the one in which 't Hooft performed
the one loop instanton computation.  The idea is that there is again an underlying
holomorphic parameter, and that the NSVZ beta function
arises by considering a suitable wave function renormalization for the
regulator fields (in particular, there must be a factor $g^{2/3}$ between these).  As we will remark, while conceptually certainly
correct, this answer is inadequate.  First, it obscures the question of scheme dependence.  Second, it is simply not
precise.  If one means by this regularization studying the theory in Wess-Zumino gauge, choosing
background field gauge for the remaining gauge field, and introducing Pauli-Villars fields for each fluctuating field, this scheme is neither BRST invariant nor supersymmetric; indeed,
it is not even a complete regulator -- higher derivative terms are required beyond one loop.
So this approach must be supplemented by finite counterterms and with them a precise renormalization scheme in every order.  It is possible to give a manifestly supersymmetric
formulation, including higher derivatives, but this scheme is still non-BRST invariant\cite{higherderivative1,higherderivative2}, and local
counterterms are still required.  In addition,
the regulated lagrangian is not
(manifestly) a holomorphic function of the gauge coupling or cutoff(s) with such a regulator, so simple arguments based on holomorphy alone are not
available\footnote{In the work of \cite{higherderivative1,higherderivative2}, the problem occurs in the action for the $B$ ghosts; it appears possible to use
holomorphy in a combinations of masses and couplings to obtain a beta function exhausted at one loop; however, because of
the other complications of this construction, particularly the finite counterterms needed
in each order to enforce BRST invarince, we will not explore these issues here; with the regulator we actually employ,
the analysis is simple and unambiguous}.  An alternative approach is based on considering theories with spontaneous breaking.  In this case, beta functions can sometimes be studied
as holomorphic functions of chiral field expectation values, leading to the NSVZ expression\cite{littlemiracles,dineshirman}.
We will explain the issue of scheme dependence in these computations
as well.

As for the instanton computation, the strong coupling computation is known to yield an incorrect result for the gaugino condensate\cite{shifmanweakcoupling,fuchs,fp}; it
disagrees with
an alternative, systematic, computation of $\langle \lambda \lambda \rangle$, which again anticipated some aspects of Seiberg's program\cite{Seiberg:1994bp,fp}.
One studies, say, a theory with gauge group $SU(N)$ and $N-1$ flavors.  If one gives each flavor a small mass, $m_f$, then one can
reliably compute in this theory, determining the vacuum state and correlation functions in that state.  The gaugino condensate is a holomorphic function
of $m$ and the coupling, and one can analytically continue the result to large $m$, where the theory is the pure supersymmetric gauge theory.
In this way, one can determine $\langle \lambda \lambda \rangle$ in a computation that is systematic at every stage; the result disagrees, by a numerical factor
($\sqrt{4/5}$ in the case of $SU(2)$) with that of NSVZ.  Additional, quite beautiful and sophisticated analyses, have verified the
validity of the weak coupling analysis\cite{hollowood,Dorey:2002ik}.  Additional discrepancies have been explored in \cite{vainshtein}.

In this paper we will provide resolutions to both of these puzzles.  For the question of the NSVZ $\beta$ function, we start with the regulator for the pure gauge
theory proposed by Arkani-Hamed and Murayama (AHM)\cite{ArkaniHamed:1997ut,ArkaniHamed:1997mj}\footnote{The recognition that the perturbed
$N=4$ theory is finite dates to the work of \cite{johansen,parkes}.}.  These authors suggested that one could regulate
this theory by starting with $N=4$ supersymmetric Yang-Mills theory (or one of the finite $N=2$ theories), and including mass terms for extra adjoint multiplets.  By carefully studying
anomalies in various rescalings, they exhibited the holomorphy
of the effective action and the NSVZ $\beta$-function.  We revisit this analysis, noting that the AHM regulator provides an {\it extremely} simple
framework in which to understand these issues.  In standard
presentations of the theory, the action is not a holomorphic function of $\tau = {8 \pi^2 \over g_0^2} + i \theta$, where $g_0$ is the ``bare"
coupling (a sensible notion in a finite theory).  A suitable rescaling allows an almost trivial
 identification of the object, $\tau$, in terms of which the tree level action {\it is} holomorphic.
The low energy effective action {\it must} be holomorphic in $\tau$, and indeed it is easy to check that the resulting effective action is holomorphic through two loops.
The AHM regulator is instructive in other respects.  It permits the definition of sensible Wilsonian renormalization group transformations, which
are manifestly holomorphic in the parameters.
The mass parameters can be treated as spurions, not only for breaking of chiral symmetries, but for conformal symmetry as well.  These remarks extend immediately
to theories which can be embedded in $N=2$ theories.  In this setup, one can easily describe, explicitly, the renormalization scheme which leads to the exact beta
function of NSVZ.  But this is readily seen to be just one of a class of renormalization schemes, each of which leads to its own ``exact" beta function.  The NSVZ scheme,
while simple, we will see is not selected by any compelling physical consideration.

The discrepancy in the strongly coupled instanton computation
has been the source of some puzzlement, and even leads to suggestions that there may be additional vacua in the theory beyond
those associated with the gaugino condensate\cite{Kovner:1997im}.  It also raises further questions about the meaning of the NSVZ $\beta$ function.
  In this note, we explain, first, that the problem already exists at the level of $\Delta$, and is not directly connected
  to the question of discrete symmetry breaking and vacuum degeneracy.
We describe why, even for short distances, there is no systematic weak coupling computation
of $\Delta(x)$.  The problem is that the theory is not under control in the infrared.  Infrared divergences if present, either in perturbation theory about
the instanton, or beyond (e.g.
in dilute gas corrections), lead to order one corrections to the leading order result, compatible with
holomorphy.  We demonstrate, using the techniques of NSVZ, that in the perturbation
expansion about the instanton, there are no such divergences.   At the same time, the dilute gas corrections are not
under control.  More precisely, {\it absent an infrared regulator, there is no argument that contributions to $\Delta$ from configurations with topological charge $1$ are exhausted
by the instanton and perturbative corrections to it}.

One needs a definition of the theory which tames the infrared; we exploit two which have
been extensively studied.
We
reconsider some aspects of the weak coupling computation \cite{fp,hollowood,Dorey:2002ik}, in which one introduces quark masses and continues holomorphic from small to large quark masses.  We focus principally on symmetry preserving Greens functions such as $\Delta$, rather than the gaugino
condensate, again observing the discrepancy.  While instructive, however, these
computations provide limited insight as to the
source of the problem.  More helpful is the theory compactified on a circle.  This theory was studied in \cite{davies}.  For small circle
radius, $\beta$ (the notation is chosen in analogy to finite temperature),
the system does admit a weak coupling description, free of infrared difficulties.   These authors were able to obtain the ground state of the system, and calculate
the gaugino condensate directly; the result agrees with the weak coupling analysis.  The calculation however, does not resemble the infinite volume
computation, involving configurations we might describe as ``half instantons."  It is interesting to calculate the correlator $\Delta$ and to understand the
origin of the discrepancy with the strong coupling result.  In the regulated theory,
the dilute gas does not contribute, but the leading instanton contribution to $\Delta(x)$ is different.  There is a contribution
which is formally identical to the conventional instanton at infinite $\beta$, but it gives a different contribution to $\Delta$, as the infinite $\beta$ limit
does not commute with the collective coordinate integral.  There are two other contributions, which do not have an $O(4)$ invariant large $\beta$ limit, yet
make finite contributions.  We identify other correlation functions, with good infrared behavior, for which the large $\beta$ limit does agree with the naive
infinite volume behavior.  These computations demonstrate unambiguously the infrared nature of the problem.   {\it They also make clear that there are contributions
to $\Delta$ in the topological charge one sector beyond those of the usual instanton solution.}

In the next section, we discuss the problem of regulators for supersymmetric gauge theories, explaining the limitations of the background
field Pauli-Villars regulator.  We explore the regulators introduced in \cite{ArkaniHamed:1997ut,ArkaniHamed:1997mj}, describing the holomorphic
renormalization group mentioned above.
In section \ref{nsvz}, we explain the exact $\beta$ function(s).    Then, in section \ref{irdivergences}, we discuss the problem of infrared divergences in strongly
coupled theories, explaining why the strong coupling calculation is not reliable.  We leave for the Appendix the demonstration of the absence of infrared divergences
in the perturbation series about the instanton, which proves a non-renormalization theorem.  In section \ref{dilutegas}, we explain why an infrared regulator is needed to define the dilute gas
corrections, which are otherwise order one and compatible with holomorphy.  In section
\ref{weakcoupling}, we discuss the weakly coupled theory, focussing on issues surrounding the computation of the correlator $\Delta$.
Then we turn to the theory on $R^3 \times S^1$ in section \ref{compactified}.  After briefly reviewing the results of  \cite{davies}, we describe
the instanton solutions which contribute to $\Delta$, and the limit of large radius.  We consider a number of issues, including the question of whether, in fact,
this is a good regulator.   Our conclusions are presented in section \ref{conclusions}.

\section{Regulating The Ultraviolet}
 \label{regulators}

Issues of holomorphy and their connections to the $\beta$ function are most easily understood in the framework of Abelian gauge theories.\cite{svabelian}
In such theories, it is enough to introduce a chiral Pauli-Villars field\cite{svabelian}.  The corresponding mass parameter is a holomorphic parameter.
If one integrates out physics between scales $m_1$ and $m_2$, the resulting effective lagrangian is\cite{dineshirman}
\beq
{\cal L}_{eff}=-{1 \over 32 \pi^2}\int d^2 \theta ~ W_{\alpha}^2 \left  ({8 \pi^2 \over g^2} + i \theta + b_0\log(m_{hol}^{(2)}/m_{hol}^{(1)}) \right ) + \int d^2 \theta ~ m_{hol}^{(2)} \Phi \Phi
+ Z(m_1,m_2) \int d^4 \theta ~\Phi^\dagger \Phi. \nonumber \\
\eeq
 where the subscript $hol$ denotes the holomorphic (bare) parameter.  Due to the requirements of holomorphy,
 evolution of the gauge couplings, in terms of these parameters, is exhausted by
 the one loop correction.  So the renormalization group equation, in terms of the holomorphic parameter, is very simple:
 \beq
 \beta_{hol} = m_{hol} {\partial g \over \partial m_{hol}} = - b_0 {g^3 \over 16 \pi^2}.
 \eeq
 However, Shifman and Vainshtein explain that it is more appropriate to use a physical mass scale, $m_{phys} = Z^{-1} m_{hol}$ in the equation, and this leads
 to an expression involving the anomalous dimension:
 \beq
 \beta =  m_{phys} {\partial g \over \partial m_{phys}} = - b_0 {g^3 \over 16 \pi^2}(1 + \gamma).
 \eeq
 We have framed the problem in terms of the Wilsonian effective action.  It can equally well be described in the language of the 1PI action, as discussed
 by Shifman and Vainshtein.  We can ask, for example, for the coefficient of $W_\alpha^2$ in this action, defined, say, in terms of the vector three point function
 at energy scale $\mu^2 \ll m^2_{phys}$.  Then, as in textbook arguments, in $\Gamma$ logs of $\mu^2$ must be logs of $\mu^2/m_{phys}^2$, so the beta functions are the
 same.  However, our definition of the ``physical" scale is arbitrary.  Even if we think of the Pauli-Villars field as physical (we will sensibly think of regulator masses
 as physical shortly), $Z^{-1} m_{hol}$ is not necessarily the location of the pole in the propagator, beyond low orders in perturbation theory.  We can redefine
 $m_{phys}$ by a function of $g^2$, correspondingly redefining the beta function beyond two loops.  This is the usual issue of scheme dependence and is equivalent to a redefinition
 of $g$.

 NSVZ first considered the non-abelian case, motivated by observations about instanton computations.   Non-abelian theories introduce new complications,
 as regularization is more complicated.
't Hooft regulated his instanton computation by introducing Pauli-Villars fields.   Often discussions of instanton computations in supersymmetric theories,
and of the NSVZ beta function, are predicated on such a regulator, with statements like ``the scheme in which the NSVZ
 beta function is exact is the Pauli-Villars scheme."  But Pauli-Villars regulators are problematic in gauge theories.  In the background field
method, as used by 't Hooft, one divides the fields into background plus fluctuation.  This formulation preserves a gauge invariance under which the fluctuating
fields transform homogeneously, and the background fields transform inhomogeneously.  This guarantees that the effective action, as a function of the background
fields, is gauge invariant.  This remains true in the presence of the regulator
  fields (one massive field for each fluctuating field).  If one works with component fields (in Wess-Zumino gauge), the regulator fields break supersymmetry
  and BRST invariance. Just as important, as is well known, they do not tame ultraviolet divergences beyond one loop, and must be supplemented by
  higher derivative regulators.  So much more is required to specify both the regulator and the renormalization scheme (remember that the goal is to make
  a statement valid to all orders in perturbation theory).  The breaking of supersymmetry can be avoided by working in a supersymmetric background field
  formalism, in which one can also introduce higher derivative regulators\cite{higherderivative1,higherderivative2}.
  But the regulator fields still spoil the BRST symmetry, which
must be enforced order by order by choice of counterterms for the interactions
of the fluctuating fields.  
Moreover, the (``bare") lagrangian is not manifestly holomorphic as a function of the gauge coupling or the regulator masses, so additional
arguments are required to account for the holomorphy of the Wilsonian action\footnote{In the formalism of \cite{higherderivative1,higherderivative2}, for example, the $B$ ghost terms violate holomorphy.  One can
write the effective action in terms of holomorphic combinations of parameters, but other subtleties remain.  Still, given our observations below,
one expects that it is possible to resolve all of these issues.}.

However, in many cases, a more satisfying regulator is available,
at least in perturbation theory\cite{ArkaniHamed:1997ut,ArkaniHamed:1997mj}.  This regulator is simple, supersymmetric, and BRST invariant.
Holomorphy, as we will see, is readily understood.
Starting with one of the {\it finite} supersymmetric theories with $N=2$ or $N=4$ supersymmetry, one can add mass terms so as to obtain regulated versions
of a large set of vector-like
supersymmetric theories.  The masses constitute {\it spurions}, not only for broken $U(1)$ symmetries,
but also for (super) conformal invariance.  Suitable quantities must be holomorphic in these mass terms.
We review these constructions, in order to apply them to the problems which interest us here.  Our point of view will be slightly different
than that of \cite{ArkaniHamed:1997ut,ArkaniHamed:1997mj}, and provides, we believe, a somewhat simpler picture.

We will focus mainly on the $N=4$ theory.  This theory has three adjoint chiral fields, $\Phi_i$, $i = 1,2,3$, and an $SU(4)$ $R$ symmetry.  If one writes the lagrangian
in a fashion so that (in terms of component fields) this symmetry is manifest,
there is a factor of $1/g^2$ in front of the kinetic terms for the chiral fields, but also a $1/g^2$ in front of the
superpotential term for the adjoints.  In a presentation in $N=1$ superspace:
\beq
{\cal L} = \int d^4 \theta {1 \over g^2} \Phi_i^\dagger \Phi_i -{1 \over 32 \pi^2} \int d^2 \theta \left ( {8 \pi^2 \over g^2} + i \theta \right ) W_\alpha^2
+ \int d^2 \theta {1 \over g^2} \Phi_1 \Phi_2 \Phi_3 + {\rm c.c.}
\eeq
In this form, the action is not manifestly holomorphic in $\tau$.\footnote{One might wish
to replace $1/g^2$ by $\lambda$ in $W$, and treat $\lambda$ as a holomorphic
parameter.  But only for $\lambda = 1/g^2$ is the theory finite, so further regularization is required
before one can discuss holomorphy in $\tau$ and $\lambda$.}  In order to exploit the power of holomorphy,
it is necessary to rescale the adjoints so that there are no factors of $g$ in the superpotential; for example
\beq
\Phi_i \rightarrow g^{2/3} \Phi_i.
\eeq
We can add mass terms for the $\Phi_i$'s (for simplicity, we will take all masses the same, but this is not necessary, and allowing them
to differ allows one to consider other questions):
\beq
{\cal L} = \int d^4 \theta {1 \over g^{2/3}} \Phi_i^\dagger \Phi_i - {{1 \over 32 \pi^2}} \int d^2 \theta \left ( {8 \pi^2 \over g^2} + i \theta \right ) W_\alpha^2
+ \int d^2 \theta (\Phi_1 \Phi_2 \Phi_3 + m_{hol} \Phi_i \Phi_i + {\rm c.c.} ).
\label{holomorphicaction}
\eeq
We will refer to this as the {\it holomorphic presentation} of the $N=4$ theory.

Working with the holomorphic presentation, holomorphic quantities calculated at low energies, and in particular the gauge couplings
in the Wilsonian action for the gauge fields (to be defined shortly)
are necessarily holomorphic functions of $\tau$ and $m_{hol}$.  Note that in this tree level lagrangian, the adjoints each have tree level mass
$g^{2/3} m_{hol}$ (this is precisely the relationship anticipated by NSVZ, eqn.(\ref{nsvzconnection})).
In the presence of the mass term, the full theory remains ultraviolet finite;   $m_{hol}$ may be thought of as a spurion,
not only for breaking of a $U(1)_R$ symmetry,  {\it but also for (super) conformal} invariance.  Holomorphic quantities must be holomorphic
functions of $m_{hol}$. Fixing their mass dependence, say, by $U(1)_R$ transformation properties, fixes their behavior under conformal rescalings.
This regulator also allows the definition of a {\it Wilsonian} action.

This setup leads to a simple understanding of the holomorphic beta function.  The main features
can be seen through simple Feynman graph and renormalization group analysis.
The simplest object to consider is the renormalization group invariant scale.  This is given by standard renormalization group arguments and anomaly considerations as:
\beq
\Lambda^{3N} = m^{3N} e^{-{8 \pi^2 \over g(m)^2}- i \theta}{1 \over g^{2N}} = (m_{hol})^{3N}e^{-{8 \pi^2 \over g(m)^2}- i \theta},
\eeq
through two loop order.  Now $m$ here is the matching scale between the low and high energy theories, which we have seen, to lowest order
in perturbation theory, is $g^{2/3} m_{hol}$.
So we see that $\Lambda$ is, to this order, a holomorphic function of $m_{hol}$.
 This must persist to all orders of perturbation theory about the instanton.

It is perhaps more instructive to use this regulator to define a Wilsonian action, and a Wilsonian renormalization group for the pure supersymmetric
gauge theory.
We can define a Wilsonian renormalization group transformation as the difference of the couplings in the 1PI actions
computed with masses $m_1$ and $m_2$, $m_1 > m_2$.  This is precisely the contribution
to the Wilsonian action obtained by integrating out physics between scales
$m_1$ and $m_2$.  For example, the coefficient of $W_\alpha^2$ is obtained by
computing a suitable 1PI coupling at a scale $\mu^2 \ll m_1^2,~ m_2^2$, including terms in the beta function through two loops, is:
\beq
{8 \pi^2 \over g^2(\mu)}= {8 \pi^2 \over g^2(m_i)} + 3N \log(\mu/m_i) - 2N \log(g(\mu)/g(m_i)).
\eeq
The left hand side must be the same for two choices of regulator mass, $m_1,m_2$.  So
\beq
{8 \pi^2 \over g^2(m_2)}  = {8 \pi^2 \over g^2(m_1)} + 3N \log(m_2/m_1) - 2 N \log(g(m_2)/g(m_1))).
\label{wilsonian}
\eeq
Now if we take
\beq
m_1 = g^{2/3}(m_1) \mh;~~~m_2 = g^{2/3}(m_2) \mmh,
\label{nsvzscheme}
\eeq
 we have, through this order, that the transformation is holomorphic as a function
of $\tau,~\mh$, and $\mmh$, i.e
\beq
{8 \pi^2 \over g^2(m_2)}  = {8 \pi^2 \over g^2(m_1)} + 3N \log(\mmh/\mh)
\label{wilsonian}
\eeq
 This must continue to hold through higher orders.

Note that we took the masses to be $g^{2/3}(m_i) \mih$; more precisely, these are the masses appearing in the propagators in perturbation theory.  In perturbation theory, $g^2$ is corrected in each order.  Correspondingly, one must introduce counterterms for $g^2$ and for the masses of the various fields.  eqn.(\ref{nsvzscheme})
is a renormalization scheme, which specifies the counterterms.  We will see that in this scheme, the NSVZ formula for the $\beta$ function holds, and the leading results for certain
instanton expressions are exact.  But one can equally well choose other schemes, and obtain different {\it exact} results.  Alternatively, one can attempt a more physically
motivated scheme, such as an ``on shell" scheme for the regulator mass.


It is worth clarifying the distinction between the Wilsonian action and the ``1PI" action.  Having defined the Wilsonian action at scale $m_2$, say,
amplitudes in the low energy theory are computed with the cutoffs $m_2$ and $g^2 m_2$, and are {not} holomorphic; e.g. as a function of momentum,
they may involve $\log(\vert m_2\vert/p^2)$.  As we will comment later, we might expect that the instanton scale parameter, $\rho$, is similar to the momentum, and so $g^2(\rho)$ factors might appear modifying the measure for the instanton collective coordinates in general renormalization schemes.   However, we will also
see that this is not critical to building an argument that $\Delta$ is not renormalized in perturbation theory about the instanton.

\section{The NSVZ $\beta$-function}
\label{nsvz}

We can reverse our earlier analysis of the holomorphy of the gauge couplings to understand the question
of exact $\beta$-functions.  Consider
 a Wilsonian renormalization group transformation involving
two $N=4$ regulator masses, $\mh$ and $\mmh$.  Then taking, at each order in perturbation theory,
the mass appearing in the propagator to be
\beq
m_i = g^{2/3} (m_i) \mih
\eeq
yields a Wilsonian renormalization group transformation:
\beq
{8 \pi^2 \over g^2(m_2)}  = {8 \pi^2 \over g^2(m_1)} + 3N \log(m_2/m_1) - 2 N \log(g(m_2)/g(m_1))).
\label{wilsonianagain}
\eeq
No further corrections can appear; this would be inconsistent with holomorphy.
Correspondingly, the $\beta$ function is that obtained by NSVZ:
\beq
\beta_{NSVZ}(g) = -{g^3 \over 16 \pi^2} {3 N \over 1 - 2N{g^2 \over 16 \pi^2}}.
\eeq
If we were to adopt a scheme
in which, for example,
\beq
m_i = g^{2/3}(m_i)f(g) \mih
\eeq
where $f(g)=1 + a{ g^2 (m_i) \over 16 \pi^2}+ \dots$,
the one loop expression would be altered, and holomorphy would necessarily imply additional corrections in eqn.(\ref{wilsonianagain}) (at three
loops and beyond).
Indeed, in this case the beta function would be given by:
\beq
\beta(g) = {\beta_{NSVZ}(g) \over 1+\beta_{NSVZ}{f^\prime \over  f}}.
\eeq
Note, for example, that for simple choices of $f$, the location of the pole (${2N g^2 \over 16 \pi^2} =1$) in the NSVZ expression moves (and can even
be eliminated).   Of course, because it occurs at strong
coupling, the significance of this pole is questionable in any case.

In some sense, the scheme of eqn.(\ref{nsvzscheme}) seems particularly well-motivated with the $N=4$ regulator.  But more careful
consideration suggests that the scheme, beyond the leading order, is rather arbitrary.  $g^{2/3} m_{hol}$ is not, in an exact sense,
a particularly significant physical scale.  It is not the location of the pole in the adjoint propagator, for example.  (Given that the theory is finite,
  we can think of the theory with particles with mass parameter $m_1$ as {\it defining} the macroscopic theory; the mass
  of the adjoint field is then a well-defined physical quantity).  For example, already at one loop, there is a {\it finite}
wave function renormalization for the $\Phi_i$ fields, which in turn corrects the propagators.  On shell one finds:
\beq
Z = 1 + \sqrt{3}\pi N{g^2 \over  16 \pi^2}.
\eeq
Such an on-shell scheme would appear physically well motivated.  We could give a prescription for determining the function $f$ in this case,
order by order in perturbation theory.

  Actually, there is a larger set of schemes, since we could reasonably introduce separate
masses for each of the $\Phi_i$ fields, and we could also do our initial rescalings so that
\beq
m_i = g^{a_i}(m_i) \mih
\eeq
where
\beq
\sum a_i = 2.
\eeq
In sum, the NSVZ scheme is mathematically simple, but not tied to any compelling physical consideration.

These observations about the NSVZ beta function, in turn, clarify certain aspects of the non-renormalization theorems for the instanton computation; after all, the
NSVZ beta function was {\it motivated} by the instanton computation.  The leading instanton result is proportional (in the case of the pure
gauge $SU(N)$ theory)  to
\beq
M_{PV}^{3N} e^{-{8 \pi^2 \over g^2}-i\theta} {1 \over g^{2N}},
\eeq
where $M_{PV}$ is the Pauli-Villars regulator mass.
NSVZ argued that this result was subject to a non-renormalization theorem; all perturbative corrections to the leading order should vanish.
If this is the case, and this quantity is renormalization group invariant, then the NSVZ form follows.
Proving the non-renormalization theorem, however, is subtle.  It does {\it not} follow from holomorphy, and we have argued that conventional Pauli-Villars regulators
must, in any case, be supplemented by additional counterterms.  On the other hand, the $N=4$ theory provides a regulator which respects all of the
symmetries, and we can reexamine this question in this context.

As we have noted, with a conventional regulator, such as supersymmetric dimensional reduction or Pauli-Villars, the question of scheme is rather obscure.  While the one
loop structure of the Pauli-Villars regulator seems simple, the regulator must be supplemented by higher derivative interactions,
as well as finite counterterms to restore BRST invariance.  All of this is much simpler in the $N=4$ deformed theory.  The regulator is supersymmetric
and holomorphic, and we have seen that a Wilsonian renormalization group analysis is readily implemented.  But we also see that while holomorphy
is readily understood, one can define a multi-parameter set of renormalization
schemes.  Perfectly supersymmetric schemes will lead to different expressions for the $\beta$ function, and in general will lead to corrections
in powers of $g(m)$ to the instanton measure (as well as in correlation functions for fixed values of the collective coordinates).

The argument for the lack of corrections to the instanton computation in perturbation theory which we provide in the appendix does not require any such
complicated considerations of non-renormalization theorems for particular schemes.  It is closer in spirit
to \cite{Seiberg:1994bp}, with the techniques of NSVZ providing reassurance that holomorphy is a reliable guide.

Finally, we should connect our analysis with that of \cite{ArkaniHamed:1997ut,ArkaniHamed:1997mj}.  To do this, we can define a more general set
of rescalings of the fields, for which
\beq
\Phi_i \rightarrow \Phi_i g^{2/3 + \alpha}.
\eeq
In order to avoid anomalous Jacobians we also rescale the vector multiplet as $V\rightarrow V g^{3\alpha}$.
In this case, the coefficient of the $\Phi$ kinetic term is
\beq
g^{-2/3+ 2 \alpha}.
\eeq
Correspondingly, writing the mass term in the superpotential as
\beq
m_\alpha \Phi_i \Phi_i,
\eeq
we have that the physical regulator mass (at tree level) is
\beq
m = g^{2/3- 2 \alpha}   m_\alpha.
\eeq
Correspondingly, $m_{\alpha}$ is related to the quantity $m_{hol}$ we have defined by
\beq
m_\alpha =g^{2 \alpha} m_{hol}.
\eeq
The coupling evolution in this scheme is also different:
\beq
{8 \pi^2 \over g^2(m^{(2)}_\alpha)} = {8 \pi^2 \over g^2(m^{(1)}_\alpha)} + 3N \log(m^{(2)}_\alpha/m^{(1)}_\alpha)  - 6N \alpha\log \left ({g(m^{(2)}_\alpha) \over g(m^{(1)}_\alpha) }\right ).
\label{nimahitoshi}
\eeq
The analysis of ref. \cite{ArkaniHamed:1997ut,ArkaniHamed:1997mj} invokes a particular choice of $\alpha$, $\alpha = 1/3$, corresponding to a canonical
kinetic term (note that this corresponds to canonical normalization for the vector multiplet).  In this case, the mass appearing in the propagator, at tree level, is $m_\alpha$.  They consider a new coupling:
\beq
{8 \pi^2 \over \hat g^2(m_\alpha)} = {8 \pi^2 \over g^2(m_{\alpha})} + 2N \log g(m_{\alpha}) 
\label{amcoupling}
\eeq
and show (their eqn.(2.10)),  that the running of $\hat g^2$ is saturated by the one loop result.  In the language we have developed
 here, this is easily understood; if we write the theory in the holomorphic presentation, with holomorphic mass $m_\alpha$, then we need
 to take $g= \hat g$ as given by equation \ref{amcoupling}.  In terms of the original coupling, $g$, we obtain the NSVZ beta function.  But we see, again, there is nothing particularly special about this
result; we can define such a coupling for {\it any} choice of $\alpha$ (as well as more general transformations).  Once more, what is critical is the holomorphy of the Wilsonian action as a function of $m_{hol}$ in eqn.(\ref{holomorphicaction}).  Again, it should be stressed that the location of the pole in the propagator at tree level is not a physical scale; in the canonical
form, this scale coincides with the mass in the lagrangian, and this has a certain elegance, but, as in the holomorphic presentation,
choosing this mass scale to define the beta function is a choice of scheme.  There are an infinity of possible schemes, each leading to an exact beta function.

\subsection{N=2}

It is interesting to contrast the $N=4$ case with the case of the finite $N=2$ theory as regulator.  Here we can consider a variety of low energy theories.
In the microscopic theory, one has a single adjoint and $N$ hypermutiplets in the $N$ representation, $N$ in the $\bar N$.
One can give mass to various combinations of the hypermultiplets, obtaining different low energy
  theories.  In this case, unlike that of $N=4$, the adjoint does not require rescaling in order to obtain
  a holomorphic form for the action.  As a result, the adjoint mass appearing in loops
  is $g^2 m_{hol}$, while those of the hypermultiplets are simply $m_{hol}^2$.  One again sees that in terms of $m_{hol}$,
  the renormalization of the coupling is saturated at one loop, and the NSVZ beta function is obtained if the cutoff is taken to be precisely
  $g^2 m_{hol}$ for the adjoints, $m_{hol}$ for the hypermultiplets.

\section{Strong Coupling and Infrared Divergences}
\label{irdivergences}

In QCD, it is not true that one can simply compute any quantity at short distances.  Indeed, this is the essence of the usual
operator product expansion analysis.  If we study a correlation function of two operators:
\beq
{\cal C}(x) =\langle {\cal O}(x)  {\cal O}(0) \rangle
\eeq
this can be replaced, as $x \rightarrow 0$, by
\beq
{\cal C}(x)=
\sum_n c_n(x) \langle {\cal O}^{(n)}(0) \rangle.
\eeq
The $c_n$'s, the coefficient functions of the operator product expansion, can be calculated systematically in asymptotically
free theories.  But the computation of the operator matrix elements, $\langle {\cal O}_n \rangle$ is, in general, a strong coupling problem.  This is familiar, not
only from perturbation theory, but
from instanton computations in QCD, which typically are infrared divergent, even for small $\vert x \vert$.  For correlators
at small distances, these divergences can be understood as arising from the computation of the operator matrix elements\cite{Dine:2010jg}.

In fact, there are quantities which are manifestly controlled by the ultraviolet, even in pure gluodynamics.  Consider, for example, the correlator
\beq
\Omega(x) = \langle {\cal O}_5(x) {\cal O}_5(0) \rangle;~{\cal O}_5 = \lambda \sigma^{\mu \nu} \lambda F_{\mu \nu}.
\label{omegaequation}
\eeq
The operator product includes terms
\beq
{g^4(x) \over \vert x \vert^4} \left(\lambda(0) \lambda(0)\right) \left( \lambda(0) \lambda(0) \right) +{ \Lambda^6 \over \vert x \vert^4} ~I.
\eeq
The first term is the most singular term in perturbation theory; the second is generated by instantons\cite{amati,Dine:2010jg}.  The second
of these references explains in detail why the one-instanton computation of the coefficient function of the unit operator
is reliable.
The matrix element of the four gaugino operator receives contributions from instantons and potentially from dilute gas corrections.
As explained in \cite{Dine:2010jg}, the dilute gas corrections do not correct the second term (there are corrections from perturbation theory
around the instanton to the coefficient, in powers of $g^2(x)$).  So for this Green's function, difficult non-perturbative corrections are suppressed
by two powers of $g^2(x)$, and a reliable computation is possible.  Necessarily, any sensible infrared regulator should obtain the same result for this
Green's function as for the theory without a regulator.  This will be an important test in what follows.

So one might say the NSVZ discrepancy is simply a strong coupling problem, and one should not be surprised.  But it would
be satisfying to have a deeper understanding.  NSVZ argued that the leading result for $\Delta$
is exact, based on reasoning about the structure of perturbation theory.  However, their arguments
require a regulator which respects certain symmetries and in addition assumes the absence of infrared divergences.
In the absence of infrared divergences, holomorphy is itself enough in any case.

The perturbation expansion of $\Delta$ about the instanton is purely formal, as there is
no small scale.  It is easy to see that divergences in higher orders,
could potentially lead to contributions which, while formally suppressed by $g^2$, are in fact of order one.  Moreover, these
would-be contributions are holomorphic.  In the next order in perturbation theory,
one might worry that there are contributions of the form
\beq
g^2 \int {d \rho \over \rho} = g^2 \log(M/\lambda).
\eeq
Here $M$ is an ultraviolet cutoff, and $\lambda$ is an infrared cutoff.  $\lambda$ is necessarily proportional to the renormalization group invariant scale,
\beq
\lambda = a \Lambda \sim a M e^{-{8 \pi^2 \over g^2}},
\eeq
so such a contribution would be of order $1$.  Because it has no $g$ dependence, it is compatible with holomorphy.

This sort of breakdown of perturbation theory is  familiar in quantum field theory in other contexts.
For example, in ordinary QCD,
the perturbation expansion of the free energy is infrared divergent, beginning at four loop order.  These divergences are believed to be cut off
by the mass scale of a three-dimensional version of QCD, with coupling $g^2 T$.  Starting at this order, the perturbation expansion is at best
formal, in powers of $g^2 T/g^2 T$.

It may seem, at first sight, farfetched that {\it holomorphic} expressions should emerge from expressions which combine holomorphic and anti-holomorphic
variables.  But this is already the case in ordinary perturbation theory.  Consider a Wess-Zumino model with a light field, $\phi$, a heavy field, $\Phi$, with
mass $m$,
and a cubic coupling $\lambda \phi^2 \Phi$.  Integrating out $\Phi$, yields a $\phi^4$ term in the superpotential; the corresponding Feynman graph behaves
as
\beq
{\lambda^2 m^* \over p^2 +\vert m \vert^2},
\eeq
and is holomorphic in $\lambda$ and $m$ as $p \rightarrow 0$.

We will argue in the appendix that it is likely that all perturbative contributions {\it do} cancel, and this is not the source of the discrepancy.
To make this claim, we first establish that perturbation theory about the instanton solution is infrared finite.  Then we invoke
holomorphy, to prove that there cannot be $g^2$-dependent corrections to the lowest order result.

\section{The Dilute Gas}
\label{dilutegas}

While perturbative corrections to the instanton do not account for the discrepancy,
it is well known that
the instanton (with all of its perturbative corrections) does not exhaust the contributions to correlation functions in the
sector with unit topological charge.  At the least, one must consider
the dilute instanton gas.  It is not possible to give a rigorous definition of these contributions. But if they are present, in the case of the pure gauge $SU(N)$ theory, they
behave as:
\beq
\Lambda^{b_0} (\Lambda^* \Lambda)^{nb_0}\int d \rho \rho^{2n b_0 -1}
\eeq
where in the case of $SU(N)$,  $b_0=3N$ and $\rho$ is shorthand for a multidimensional integral over approximate moduli rescaling the different
instantons and anti-instantons.  Assuming that the infrared divergences are cut off at a scale
$\rho \sim \lambda^{-1}$, the result is proportional to:
\beq
\Lambda^{b_0}{ (\Lambda^* \Lambda)^{n b_0} \over \lambda^{(2n) b_0}},
\eeq
i.e. it is of the same form, again, as the leading order result.

The sorts of arguments we have made in the appendix for perturbation theory will not apply to these configurations.  Consider, say, a configuration
 with two instantons and an anti-instanton.  This configuration is not a solution of the equation
  of motion, and breaks all of the supersymmetries and superconformal symmetries.  When these objects approach one another, the various fermion zero modes will be lifted.  In the strongly coupled theory, it is not clear that there is any way to perform this computation in a systematic fashion.  It is hard to see a reason of principle that such contributions should vanish, and, as we have seen, {\it any} non-vanishing correction would represent an order one contribution
  to the leading instanton result.

In a strongly coupled theory, of course, isolating and defining the dilute gas contributions is problematic.  But this very issue indicates that there is no reason to trust the
leading semiclassical analysis.  Stated more generally, one
expects that the contribution to $\Delta$ of the unit topological charge sector is not
exhausted by the standard single instanton solution.
One might expect that if one has a procedure to define the dilute gas contributions, one will find a contribution of order one; alternatively, the dilute
gas contributions might vanish, but because of the infrared sensitivity of the theory, the single instanton contribution might be modified from its naive value.  In the next
two sections, we will consider two infrared regulators.  In both cases, because the theory is well-behaved in the infrared, holomorphy
{\it does} mean that there are no dilute gas contributions.  In the ``weak coupling calculation" (small quark mass, analytically continued to large quark
mass), the origin of the failure of the strong coupling calculation is obscure.  In the case of the theory compactified on a circle,
one can see more directly the source of the problem, as one can identify, for large compactification radius, $\beta$, the would-be strongly-coupled instanton, and the calculation
is under control both in the small and large $\beta$ regime.

\section{Regulating the Infrared}

The discussion of the previous section demonstrates that one cannot assess the validity of the NSVZ calculation without a suitable infrared regulator (we have already
encountered the problem of ultraviolet regularization).  There are two such regulators which have been considered in the literature.  One, due to Seiberg,
involves first studying a supersymmetric gauge theory with additional quark fields with small masses.  In this theory, weak coupling computations are possible.
Using holomorphy and symmetries, one can then continue to large mass.   While this argument is usually used to compute the gaugino condensate, it is also possible,
as we explain here, to compute Green's functions like $\Delta(x)$, but not $\Omega(x)$ (eqn.(\ref{omegaequation})).  A second regulator is achieved by compactifying
the Euclidean time direction on a circle of radius $\beta$, with periodic boundary conditions\cite{davies, Diakonov:2002qw}.  This system has been studied in the limit of small $\beta$, where it has been shown that one obtains the correct value of the gaugino condensate.  But it is possible to study, as well,
Green's functions like $\Delta$ {\it and} $\Omega$, and determine whether they agree with the weak coupling result, and if so the origin of the discrepancies with the strong
coupling computation.  We consider both of these questions in the next subsections.

\subsection{Instantons in the Weak Coupling Theory}
\label{weakcoupling}

Before continuing further with the strong coupling problem, it is useful to review some aspects of instanton effects in weak coupling theories.
There is a large literature on this subject, beginning with\cite{shifmanweakcoupling}; the main points are thoroughly reviewed and developed in \cite{hollowood,Dorey:2002ik}.
Our main purpose here is to point out that arguments close to those of NSVZ are applicable to
these systems, and, because they admit a systematic weak coupling analysis, they can be readily tested.
Then we discuss the correlator $\Delta(x)$, explaining that, again, it must be a constant, and that, as argued by NSVZ, it must factorize for large $x$
as a power of the gaugino condensate.  We stress that, as a result, the issue in NSVZ is not merely the question of factorization,
but {\it the value of $\Delta$ itself}.

To compute the gaugino condensate\cite{Seiberg:1994bp}, one works in a theory with, say, gauge group $SU(N)$ and $N_f$ quark flavors, $Q_f$ and $\bar Q_f$.
One adds a mass term:
\beq
W = m_{hol} \bar Q_f Q_f
\eeq
(to simplify the writing we will take all of the masses the same, but this is not necessary).
At low energies (well below the cutoff, $M$), the mass is renormalized due to the wave function renormalizations of the $Q$'s:
\beq
K = Z Q_f Q_f^\dagger;~~~Z = \left ( {g(m_{phys}) \over g(M)} \right )^{-2{N^2 -1 \over N b_0}},
\eeq
where $b_0=3N-N_f$.
The physical and holomorphic masses are related by
\beq
m_{phys} = Z^{-1} m_{hol}.
\eeq
For small mass, the quarks have large
vev's and the system is weakly coupled.  In addition, the discrete $R$ symmetry of the system
is spontaneously broken.  One can calculate a variety of gauge invariant correlation functions.  These include
the gaugino condensate, as well as correlation functions with more gauginos and with scalar fields.  Arguments
based on holomorphy are reliable in these models (and have been verified in some cases\cite{hollowood,Dorey:2002ik}).

There are two related issues with the NSVZ program which are realized and readily understood in weakly coupled situations.
The first is the extent to which the result of the computation is not renormalized, and the second is the meaning of the {\it exact (NSVZ) $\beta$ function}.

To address these, it is interesting to do the following exercise, essentially a minor extension of the analysis in \cite{shifmanweakcoupling,Seiberg:1994bp}.  We take
the large mass limit of the weak coupling result for $\langle \lambda \lambda \rangle$, and check matching to the next to leading order.
This is a somewhat non-trivial check of holomorphy, as it is crucial that there be no factors of $\log(\vert m_{phys} \vert)$ in the expression;
these would arise from $g^2(\vert m_{phys} \vert)$ in a renormalization group analysis.

The basic idea of \cite{shifmanweakcoupling,Seiberg:1994bp} is to note that in the low energy theory,
\beq
\langle \lambda \lambda \rangle = \Lambda_{LE}^3.
\eeq
From the renormalization group,
\beq
\Lambda_{LE}^3 = m_{phys}^3 e^{-{8 \pi^2 \over N g(m_{phys})^2}}{1 \over g(m_{phys})^{2}}.
\label{matching}
\eeq
Now one writes:
\beq
{8 \pi^2 \over g^2(m_{phys})} = {8 \pi^2 \over g^2(M)} + b_0 \log(m_{phys}/M) - {b_1 \over b_0} \log \left ({g(m_{phys}) \over g(M) }\right ),
\eeq
where $b_0=3N-N_f$ as before and $b_1 = 2N b_0 - 2\frac{N_f}{N} \left(N^2-1\right)$.
So one immediately finds:
\beq
\langle \lambda \lambda \rangle = (m_{phys})^{N_f/N} (\Lambda)^{3N-N_f \over N}
\label{LowHighMatching}
\eeq
with all quantities evaluated through second order.
The scale $\Lambda$ is defined by
\begin{equation}
\Lambda^{b_0} = M^{b_0} e^{-\frac{8\pi^2}{g(M)^2}} \frac{1}{g(M)^{2N}}.
\end{equation}
There are no factors of $g(\vert m_{phys} \vert)$ on the right hand side of eqn.(\ref{LowHighMatching}); indeed, the structure of the
result is fully determined by symmetries and holomorphy.  This
must persist through higher order.  We naturally identify the coefficient of $m_{phys}^{N_f/N}$ with the {\it holomorphic} $\Lambda$.

Let's turn to the question of correlators and factorization.
Consider correlators of the form
\beq
G_k(x_1,...x_k) = \langle \left( \lambda(x_1)\lambda(x_1) \right) \dots \left( \lambda(x_k) \lambda(x_k) \right) \rangle.
\eeq
We can first ask what sorts of instanton configurations contribute to $G_k$.  It is helpful, here, to consider the anomalous $U(1)$ under which
all of the scalar fields are neutral.  It is necessary to assign $\Lambda$ charge $2/(3N-N_f)$.  Noting that the $n_I$ instanton
amplitude is proportional to $\Lambda^{b_0 n_I}$, it follows that $G_k$, which carries charge $2k$,  gets contributions from $k$ instanton
configurations.

As in the strongly coupled case, $G_k$ must, in fact, be independent of position.  In the weakly coupled theory, it can be evaluated
by taking the coordinates far apart.  Since the gauge symmetry is completely broken, almost all propagators fall to zero exponentially
with distance, and the $k$ instanton calculation factorizes into a product of $k$ single instanton contributions.
In other words, in the weak coupling theory,
\beq
G_k(x_1,\dots,x_k) = \langle \lambda \lambda \rangle^k.
\eeq
This is precisely the factorization discussed by NSVZ.  Now, however, the discrete symmetry is spontaneously broken from the start;
there is no sum over different vacua.
Note also that it is clear that the discrepancy between the strongly coupled and weakly coupled instanton computations is {\it not} a consequence
of factorization; it holds already at the level of the full Green's function.

\subsection{The Compactified Theory}
\label{compactified}

An alternative regulator is provided by the compactification
of the theory on $R^3 \times S^1$\cite{davies}.  This is essentially a three dimensional gauge theory with scalars ($A_4$) in the adjoint
representation.
The theory has a classical flat direction, in the case of $SU(2)$,  in which
\beq
A_4 = v {\sigma_3 \over 2}.
\eeq
Dualizing the gauge field allows one to write the theory in terms of a theory with four supersymmetries, and a chiral field, $\Phi$, whose scalar components consist of
$v(x)$ and the dual of the gauge field.
In this theory, the simplest instantons are (from a four dimensional perspective) static magnetic monopoles.
The authors of \cite{davies} showed that monopoles generate a superpotential for $\Phi$.
In the case of $SU(2)$, there are actually two types of monopoles, the usual BPS monopole, and a transformation of the monopole by
\beq
U = e^{-i{ \pi x_4 \over \beta} \sigma_3}.
\label{monopoletransformation}
\eeq
This is referred to as the ``KK monopole".
The potential has two minima, corresponding to
breaking the $Z_2$ symmetry.  In each vacuum, one can calculate $\left< \lambda \lambda \right>$ by summing the
contributions from these two monopoles (each of which has two zero
modes),
\beq
\langle \lambda \lambda \rangle = \langle \lambda \lambda \rangle _{BPS} + \langle \lambda \lambda \rangle_{KK},
\eeq
remarkably obtaining the weak coupling result.  As a consequence of holomorphy, the leading result is exact if expressed in terms
of the holomorphic coupling and cutoff.  However, various parts of the computation can (and do\cite{betadependence}) receive corrections in powers of $g^2$.
In principle, these can depend on $\beta v$ (a constant in the vacuum), and $\beta M$, where $M$ is the cutoff; these corrections cancel in the total condensate.

But to understand the failure of the strong coupling calculation, it is more interesting to consider the computation of
\beq
\Delta(x) = \langle \left( \lambda(x) \lambda(x) \right) \left( \lambda(0) \lambda(0) \right) \rangle.
\eeq
This will be generated by instantons which can be described as
monopole pairs.  We can label the three types of solution as $BPS-BPS$, $BPS-KK$, and $KK-KK$.
Consider, first, the limit $\vert x \vert \gg \beta$.  Then $\Delta$ will be generated by configurations where one
monopole is near $x$ and one near $0$.  Clearly for large separations we have
\beq
\Delta = \langle (\lambda \lambda)( \lambda \lambda) \rangle_{BPS-BPS}
+  \langle (\lambda \lambda )(\lambda \lambda) \rangle_{BPS-KK}+  \langle (\lambda \lambda)( \lambda \lambda) \rangle_{KK-KK}
\eeq
$$~~~~=
\langle \lambda \lambda \rangle \langle \lambda \lambda \rangle.
$$

To compare with the strong coupling calculation, we are interested in the opposite limit, $\vert x \vert \ll \beta$.
Note that we can take this limit while keeping $\beta \ll \Lambda^{-1}$, so the coupling is still weak.
Because the correlation function is a constant in $x$, the result obtained by the simple, factorized computation still holds
in this other limit.

This raises a puzzle, but also provides part of the resolution of the instanton paradox. Consider the three
types of instanton more carefully.  The BPS-KK instanton is known explicitly from finite temperature studies\cite{leeinstanton, Kraan:1998kp}.  If one takes the limit
$\beta \rightarrow \infty$, this solution reduces to the infinite volume solution.  The collective coordinates are the six coordinates describing the location of the monopoles, as well as two $U(1)$'s.   As $\beta \rightarrow \infty$, this
solution maps to the usual instanton \cite{leeinstanton, Kraan:1998kp}.
The $\rho$ collective coordinate (up to a factor of $\beta$) measures the separation of the monopoles; the position is a linear combination of
the average of the location collective coordinates, plus a location in $x_4$.  The remaining coordinates map into the rotational collective coordinates.

Much is known about the BPS-BPS solution \cite{Weinberg:2006rq}, though an analytic form is not available. Nevertheless it has some obvious features, relevant to the questions we are addressing here.
Most important, it is independent of $x^4$.  So it cannot lead, in general, to $O(4)$ invariant expressions for Green's functions.  On the other hand,
its contribution to $\langle \lambda \lambda \rangle$ is a constant.  So this is a contribution to $\langle \lambda\lambda \rangle$ which survives
in the large $\beta$ limit.  At the same time, this instanton must not contribute in this limit to correlation functions like $\Omega(x)$ (eqn.(\ref{omegaequation})).
We will see in the next subsection that this is the case.  The $KK-KK$ instanton is obtained by performing a transformation on the gauge field
by the transformation function $U$ of eqn.(\ref{monopoletransformation}).  While this is not precisely a gauge transformation, the gauge invariant Green's functions
under consideration (Green's functions of local operators, not involving Wilson lines) are unchanged, so the $KK-KK$ instanton contributions are identical to the
$BPS-BPS$ contributions.

The BPS-KK instanton, while formally identical to the infinite volume theory at large $\beta$, also makes a different contribution than that found
by NSVZ.  This is because the limit $\beta \rightarrow \infty$ does not commute with the collective coordinate integrals.  Simple dimensional analysis indicates how this can happen.
One can expand the integrand in powers of $1/\beta$, for small $x$.  However, upon integration, simply on dimensional
 grounds, the result, if non-zero, must be (infrared) divergent, cut off at $x \sim \beta$.

 We can summarize as follows.  If one works in the formal infinite volume limit, there is no systematic computation of $\Delta$; the infrared is not under control, and
 there is no approximation scheme.  In a situation where systematic computations are possible, an infrared cutoff is present.  There are contributions to $\Delta$ which survive in the
 limit the cutoff is taken to infinity,  So the formal infinite volume result has, as expected, infrared sensitivity.

\subsection{Aspects of the Instanton Computation in the Compactified Theory}

The first question one must ask is in what sense compactifying the theory on a circle provides an infrared regulator.  Gauge theories at high temperatures exhibit
infrared divergences in perturbation theory connected with the fact that the modes with zero Matsubara frequency are governed by a three dimensional field
theory.   Feynman diagrams for the free energy, for example, exhibit infrared divergences at four loop order.  Feynman diagrams for (gauge invariant) correlation
functions of bosonic operators can exhibit infrared divergences, typically, again, at four loop order.  There is no simple argument that these divergences should cancel.  Instead,
it is generally believed that the divergences are cut off by the mass-scale of the three dimensional theory.  With periodic boundary conditions
for the fermions, new types of divergences can appear.  One might anticipate cancelations in certain correlators (e.g. lowest components of chiral fields).  But in general, one does
not expect that such a regulator completely resolves the problems of the infrared.

That said, it is also well-known that the leading terms in the finite temperature instanton computation in QCD {\it are} infrared finite.  While to our knowledge the infrared behavior
of the perturbation expansion about the instanton has not been carefully studied, we might expect infrared divergences to appear at some order as well.  Our observations
below (and those of ref. \cite{davies}), presume that, for certain quantities which are protected by supersymmetry and holomorphy, these divergences cancel.
But this is a question worthy of further study.

With this caveat, we consider some aspects of the instanton solution in the compactified theory.  First, we prove that each type of instanton contributes
a constant to the Green's function $\Delta(x)$ (i.e. the contribution of each type of instanton is independent of $x$).  
We mostly follow the standard argument, but we consider the constancy of the contribution from each configuration.
A similar argument is given in \cite{davies}.

Note that there are three types of configurations: $BPS - BPS$, $BPS-KK$ and $KK-KK$.
Let us choose one of them and let $d\mu$ denote the corresponding measure for the collective coordinates.
Then we need to calculate
\begin{equation}\label{eq:dlambda4}
\frac{\partial}{\partial{x^{\alpha\dot\alpha}}} \int d\mu\,  \lambda\lambda(x) \lambda\lambda(0)
\sim \int d\mu\, \left[ \{ Q_{\alpha}, \bar Q_{\dot\alpha} \}  \lambda\lambda(x) \right] \lambda\lambda(0)
\end{equation}
Using the fact that $\left[ \bar Q_{\dot\alpha},  \lambda\lambda(x)\right]=0$, the integrand becomes
\begin{eqnarray}
\left[ \{ Q_{\alpha}, \bar Q_{\dot\alpha} \},  \lambda\lambda(x) \right] \lambda\lambda(0)
&=&\{ \bar Q_{\dot \alpha},\left[ Q_{\alpha},  \lambda\lambda(x) \right] \} \lambda\lambda(0)\\
&=&\{ \bar Q_{\dot \alpha},\left[ Q_{\alpha},  \lambda\lambda(x) \right] \lambda\lambda(0) \} \nonumber.
\end{eqnarray}
Since the measure $d\mu$ is invariant under supersymmetry transformations, eqn.(\ref{eq:dlambda4}) vanishes, which implies that the contribution of each gauge field configuration to the correlation $\left< \lambda\lambda(x) \lambda\lambda(0) \right>$ is independent of $x$.

We now ask why the function $\Delta(x)$ can receive contributions at large $\beta$ from the $BPS-BPS$ and $KK-KK$ instantons, while the function
$\Omega(x)$ does not.  Here the question is where the support of the zero modes and the classical solution lie.  While the $BPS-BPS$ solution
is independent of $x_4$, the instanton and its associated zero modes have support on scale $\beta$.  So as $\beta \rightarrow \infty$, these functions fall
off, at any given point, as $1/\sqrt{\beta}$.  So additional fields, as in eqn.(\ref{omegaequation}), result in suppression by powers of $1/\beta$.

\section{Conclusions}
\label{conclusions}

Supersymmetry has provided a tool with which to obtain a range of exact results in field theory and string theory.  Arguably the first inkling that one
could obtain such results was the work of NSVZ.  They argued for two exact results in gauge theories.  First, that one could compute certain correlation
functions exactly at weak coupling, and extend the results to strong coupling; second, that one could obtain exact expressions for $\beta$-functions.
However, each of these results raised questions.  As methods exploiting systematic weak coupling expansions and holomorphy were developed, it became
clear that the strong coupling instanton computation was incorrect.  This in turn called the exact $\beta$-function into question; even without this issue,
the meaning of the NSVZ beta function was obscure.  The $\beta$ function is scheme dependent; in what scheme did the result hold?

In this paper, building on earlier work, we have provided answers to both questions.  For the NSVZ beta function, we have revisited the regulator of AHM.  We have
seen, first, that if the theory is presented in a fashion which is manifestly holomorphic in the gauge coupling and the regulator mass, the gauge coupling
is manifestly holomorphic.  Moreover, in this setup, one can define a Wilsonian renormalization group and Wilsonian action, both of which
are holomorphic, and exhausted, as a result, by one loop corrections (plus possible non-perturbative corrections).

For the problem of the instanton computation, we have noted, first, that infrared divergences spoil arguments based on holomorphy for non-renormalization.  Such
corrections, if present, either perturbatively or non-perturbatively, correct the leading term by effects of order one.  Thus it is necessary to perform the
computations in a fashion that the infrared is regulated.  Perhaps the most useful such regulator is provided by compactification of one direction on a circle of
radius $\beta$.  Weak coupling, semiclassical calculations can be performed both for small {\it and} large $\beta$ (relative to the length scales in the
correlation functions).  One can identify the instantons for this system, and one can consider a limit which resembles the infinite volume theory in which the system is
weakly coupled.  One can then see that the large $\beta$ limit of the theory is singular, reflecting the underlying infrared sensitivity of the system.  Indeed,
we have seen that there are three types of instanton.  For one, taking the limit of infinite $\beta$ before performing the collective coordinate integrals
leads to an incorrect result.  The other two solutions have vanishing support for finite $\vert x \vert$ as $\beta \rightarrow \infty$, yet they make
contributions to $\Delta(x)$ which survive in the limit.  As we explained
in the introduction, all of this is as one should have expected.  The fact that some subset of corrections happen to vanish does not assure that
the leading computation is reliable.  Perhaps this is the most important object lesson of this story:  calculations in situations where there is no systematic
approximation, however pretty, can be terribly misleading.

\noindent
{\bf  Acknowledgements:}
M.~Dine thanks Stanford University and the Stanford Institute for Theoretical
Physics for a visiting faculty appointment while some of this work was performed; he also thanks the Institute for Advanced
Study for its hospitality.   We thank Nathan Seiberg, for conversations which stimulated this work and for valuable criticism, and
Tom Banks, John Kehayias and Zohar Komargodski.  We thank Misha Shifman and Arkady Vainshtein for invaluable conversations explaining
their work, and for a critique of an early version of this manuscript.  We are very appreciative of
conversations with Nima Arkani-Hamed and with David Shih, who provided us with his (as yet unpublished) notes on superconformal
symmetry.  We thank Nick Dorey for explanations of aspects of the calculation in the compactified theory.
This work was supported in part by the U.S.~Department of Energy. G.~Festuccia is supported by a grant-in-aid from the National Science Foundation: grant number NSF PHY-0969448.

\appendix
\section{Appendix:  Non-Renormalization Theorem for Perturbation Theory}
\label{nrtheorem}

In this appendix, we ask whether, in perturbation theory about the instanton of the strongly coupled theory (the pure gauge
theory), there might be terms of the type
\beq
g^2 \int {d \rho \over \rho}
\eeq
or, more generally,
\beq
g^{2n} \int {d\rho \over \rho} \ln^{n-1}(\rho M)
\eeq
spoiling the perturbation expansion and giving rise to order one corrections to the Green's function.

We will argue that there is a non-renormalization theorem.  The argument has two parts.  First, we demonstrate that there are no infrared divergences in
perturbation theory.  This argument is closely tied to the
collective coordinate discussion of NSVZ.  Having established the absence of such divergences, we can then invoke holomorphy to
argue for the absence of corrections.

 First, we need to understand the nature of the collective coordinates and the collective
coordinate measure.  NSVZ\cite{Shifman:1999mv} used symmetry arguments to assert that it is possible to write an exact expression
for this measure.  We will argue that, at best, these statements are scheme dependent.  On the other hand,
using the techniques of NSVZ, it is quite simple to show that there are no infrared divergences
in the perturbation theory about the instanton.

At the classical level, the collective coordinates include symmetries, conformal and
superconformal symmetries, which are {\it not} symmetries of the
quantum theory.  We have seen that with AHM regulator, we can treat the regulator mass systematically as a spurion, not only
for conventional chiral symmetries, but for the full superconformal group ; allowing
for transformations of the spurion, the theory is invariant.  This allows us
to describe possible modifications of the measure in a systematic fashion, and to understand
the implications of the symmetry for Green's functions.

\subsection{Collective Coordinates for the Instanton Computation}

The low energy theory, classically, possesses Poincare and superconformal
symmetries.  The instanton breaks conformal invariance, translations, and half of the supersymmetry and superconformal transformations.  Correspondingly, one has
collective coordinates:
\begin{enumerate}
\item  translations, $P^\mu \leftrightarrow x_0$.
\item  scale transformations:  $L \leftrightarrow \rho$.
\item  supersymmetry transformations, $Q_\alpha \leftrightarrow (\theta_0)_\alpha$.
\item  superconformal transformations, $\bar S_{\dot \alpha} \leftrightarrow (\bar \beta)_{\dot \alpha}$.
\end{enumerate}

We are interested in the collective coordinate measure, and the dependence of Green's functions (for fixed values
of the collective coordinates) on the collective coordinates.  If both of these separately
are unchanged in perturbation theory about the instanton, clearly there is no correction, but the actual requirement is weaker.  Indeed, from our discussion
 of scheme dependence, we see that at the very least any statement about the exactness of the measure or the Green's function is contingent on a renormalization scheme.
We will see shortly that, because there are no infrared difficulties in the computation, holomorphy is enough to insure the
absence of corrections to the final result.  So we will confine ourselves to a few remarks about the measure.

In the low energy theory.
the $R$ symmetry and the conformal invariance are anomalous.  In the full theory, these symmetries are restored provided one properly transforms the
spurion.
The measure:
\beq
d^4 x_0 d^2 \theta_0 d^2 \bar \beta {d\rho \over \rho} (\Lambda \rho)^{b_0}.
\label{measurefactor}
\eeq
respects the full set of symmetries, where $\Lambda$ is the holomorphic scale:
\beq
\Lambda^{b_0} = m^{b_0} e^{-\tau}.
\eeq
Exactness of the measure
does not  follow simply from conformal invariance; if the measure were multiplied by
\beq
f(g(\rho)) = 1 + a {g^2(\rho) \over 16 \pi^2}+ \dots,
\eeq
it would still respect the conformal symmetry.  Such terms, and similar terms in the measure, would have the potential to
introduce infrared divergences of the type we have discussed.
NSVZ provided arguments based on the structure of the Feynman graph expansion, that eqn.(\ref{measurefactor}) is exact.  Our arguments
below provide support for the idea that this statement holds, at least in some suitable renormalization scheme.
But we will see this is not of great importance.




\subsection{Perturbation Theory Around the Instanton}

The procedure for computing correlation
functions can then be summarized simply.   Again, we specialize to $SU(2)$ and study the correlation function
\beq
\Delta(x_1,x_2,\theta_1,\theta_2)=
\langle W_\alpha^2(x_1,\theta_1) W_\beta^2(x_2,\theta_2) \rangle.
\eeq
While we are ultimately interested in the lowest component of $\Delta$, for fixed values of the collective coordinates,
higher components will be of interest.
First, note all components of this Green's function vanish in ordinary perturbation theory.  This follows from
\begin{enumerate}
\item  The general form of such chiral Green's functions\cite{Wess:1992cp}.
\item   The chirality properties of the operator (i.e. the fact that the Green's function carries charge $4$ under the anomalous $U(1)$ symmetry).
\end{enumerate}
In a component calculation, the perturbative vanishing of the higher components of the correlation function results from cancelations between the
$F_{\mu \nu}^2$ and $F \tilde F$ correlators, and the $\lambda \partial_\mu \sigma^{\mu *}\lambda^*$ correlators.

In order to compute the Green's function of interest (the lowest component of $\Delta$), we can, following NSVZ, proceed as follows:
\begin{enumerate}
\item  Compute  $\Delta$ for $\theta_0=\bar \beta=0$, general $\rho$ and instanton coordinate $0$.   While in principal one wants the general
expression, it is in fact only necessary to compute the $\theta_1^2 \theta_2^2$ component of $\Delta$.
\item  Act on the result with
\beq
e^{i P^\mu x^{\mu}_0} e^{i \theta_0 Q} e^{i \bar \beta \bar S}.
\label{collective}
\eeq
\item  Integrate over the result with the measure of the previous section, possibly with corrections in powers of $g^2(\rho)$.

The operation of step (2) can be implemented, alternatively,
by noting that the final result must be supersymmetry invariant.  Examining the supersymmetry transformation properties of the collective coordinates,
one sees that one should make the replacements, in the correlators of point (1) above:
\beq
\theta \rightarrow \theta - \theta_0 - (x^\mu - x_0^\mu )\sigma^\mu \bar \beta.
\label{replacements}
\eeq
\end{enumerate}

Let's review how this works for the leading order instanton computation.
We have, in the case of $SU(2)$:
\beq
W_{\alpha}^2(x_1,\theta_1) = \theta_1^2 {\rho^4 \over ((x_1 - x_0)^2 + \rho^2)^4}.
\eeq
Making the replacements of eqn.(\ref{replacements}),
\beq
\int d\rho d^4 x_0  d^2 \theta_0 d^2 \bar \beta \langle W_{\alpha}^2(x_1,\theta_1) W_{\beta}^2(x_2,\theta_2)\rangle
\eeq
$$~~~~= \int d^4 x_0  {d\rho \over \rho^5} ( \Lambda \rho)^6(x_1 - x_2)^2
{\rho^4 \over ((x_1 - x_0)^2 + \rho^2)^4}{\rho^4 \over ((x_2 - x_0)^2 + \rho^2)^4}.
$$
The $x_0$ integral is straightforward (it can be performed by introducing Feynman parameters, for example), and yields
\beq
\int {d\rho}d\alpha ~\alpha^3~ (1-\alpha)^3  ( \Lambda \rho)^6{(x_1 - x_2)^2 \rho^3 \over ((x_1 - x_2)^2 \alpha(1-\alpha) + \rho^2)^6}.
\eeq

Certain features of this result should be noticed.  One might have expected, from experience with correlation functions involving fermions in the fundamental representation,
that the expression would be infrared divergent; on dimensional grounds, one might have expected to encounter $\int d\rho/\rho$.  This does not happen because of the
factors of $(x_1 -x_2)^2$ appearing in the numerator of this expression.  As we now explain, this is general; as a consequence of symmetries,
the correlation function vanishes as $x_1 \rightarrow x_2$.  This
guarantees the good infrared (large $\rho$) behavior of the correlation
function.  The absence of infrared divergences permits the use holomorphy to establish the non-renormalization theorem, since,
in the absence of infrared divergences, corrections in (inverse powers of) $\tau$ are forbidden.

To see that this feature persists in perturbation theory about the instanton, we first
 understand the vanishing of the Green's function as $x_1 \rightarrow x_2$ another way.  Acting as in eqn.(\ref{collective}), terms quadratic in $\theta_0$ and
$\bar \beta$ can only arise from the $\theta_1^2\theta_2^2$ in the Green's functions  (essentially the $\theta_1^2(F(x_1)^2 + \dots) \theta_2^2(F(x_2)^2 + \dots)$ terms).
A simple computation shows that this operation is proportional to $(x_1 - x_2)^2$.  So provided the Green's function is not singular as $x_1 \rightarrow x_2$,
there are no infrared divergences.

To generalize to higher orders, we need to investigate possible corrections 
to the form (for fixed values of the collective coordinates) of the Green's function as $x_1 \rightarrow x_2$.  We will discuss this issue in the next
subsection, where we will argue that the Green's functions are non-singular as $x_1 \rightarrow x_2$.

\subsection{Behavior of Green's Functions in the Instanton Background}

From the arguments of NSVZ, one can construct the desired correlation function if one knows
the $\theta_1^2 \theta_2^2$ behavior of the correlation function
\beq
\langle W_\alpha^2(x_1, \theta_1) W_\beta^2(x_2,\theta_2)\rangle.
\eeq
In ordinary perturbation theory, this Green's function vanishes as a consequence of fermi-bose cancelations.  This follows from standard arguments for the structure
of chiral Green's functions.
\beq
\langle \Phi(x_1,\theta_1) \Phi(x_2,\theta_2) \rangle = \langle \Phi(x_1, \theta_1 - \theta_2, \bar \theta_1) \Phi(x_2, 0, \bar \theta_2) \rangle.
\eeq
But this correlator vanishes in perturbation theory as a consequence of the chiral symmetry of perturbation theory.

About the instanton,
at lowest order, this is just $F^{cl~2}_{\mu \nu}(x_1) F^{cl~2}_{\rho \sigma}(x_2)$.  We saw that it is crucial that this is
non-singular as $x_1 \rightarrow x_2$; then the Green's function of interest, obtained from the NSVZ zero mode argument, vanishes as $x_1 \rightarrow x_2$, which is what we need
to insure the absence of infrared divergences.

At higher orders, we again work with {\it component fields}, calculating the correlator of the $\theta_1^2$, $\theta_2^2$ operators\footnote{Note that there are actually two rotationally
invariant combinations, but we may take the second to be $(\theta_1 - \theta_2)^2$, which does not contribute.}
This is some (in general complicated, non-local) function of $x_1,x_2$.  As before, to establish whether there is an infrared singularity in the $\rho$ integral,
we can study the limit $x_1 \rightarrow 0,x_2 \rightarrow 0$.  In this case, once we make the replacement of eqn.(\ref{replacements}); the
coefficient of $\alpha^2 \bar \beta^2$ vanishes.

But we might still worry that correlation functions might be singular as $x_1 \rightarrow x_2$; after all, perturbative Green's functions are singular.
The absence of singularities, however,
is plausible; for small $x_1-x_2$, the background instanton should be unimportant, and the propagators should reduce to those of the trivial background.  For these,
 as we have explained, there is a cancelation between fermions and bosons.  One can develop a perturbation expansion, for short distances,
 of the propagators in terms of free propagators, and show that the corrections are non-singular.  As a check, one can consider known propagators.  In particular,
ref.  \cite{Creamer:1977vf} provides an expression for the isospin one scalar propagator in an $SU(2)$ instanton background.
 Taking the limit $x_1 = (1 + \epsilon) x_2$, for example, one finds that there is a $1/\epsilon^2$ singularity, which is just the free field singularity, and the term
 of order $1/\epsilon$ cancels.


Now it is also crucial that {\it logarithmic} modification of the measure, as contemplated in the previous section, even if present, will not alter this.  Factors of $\log(\rho \vert m \vert)$
would not render the integrals infrared divergent.

We can now prove the non-renormalization theorem simply.
In the absence of infrared divergences, holomorphy for the Green's function must hold order by order in the coupling expansion for $g^2$.
It may be possible to make this proof more direct, but it is adequate to establish the main result.
To all orders in the perturbation expansion about the instanton, there are no corrections to the lowest component of $\Delta$.  The discrepancy between the
strong and weak coupling computations must be understood differently.

\bibliography{susyinstantonrefs}{}

\begin{thebibliography}{10}

\bibitem{Novikov:1983ee}
V.~A. Novikov, Mikhail~A. Shifman, A.~I. Vainshtein, and Valentin~I. Zakharov.
\newblock {Instanton Effects in Supersymmetric Theories}.
\newblock {\em Nucl. Phys.}, B229:407, 1983.

\bibitem{Novikov:1983uc}
V.~A. Novikov, Mikhail~A. Shifman, A.~I. Vainshtein, and Valentin~I. Zakharov.
\newblock {Exact Gell-Mann-Low Function of Supersymmetric Yang-Mills Theories
  from Instanton Calculus}.
\newblock {\em Nucl. Phys.}, B229:381, 1983.

\bibitem{Shifman:1999mv}
Mikhail~A. Shifman and Arkady~I. Vainshtein.
\newblock {Instantons versus supersymmetry: Fifteen years later}.
\newblock 1999.

\bibitem{amatireview}
D.~Amati, K.~Konishi, Y.~Meurice, G.~C. Rossi, and G.~Veneziano.
\newblock {Nonperturbative Aspects in Supersymmetric Gauge Theories}.
\newblock {\em Phys. Rept.}, 162:169--248, 1988.

\bibitem{exactbeta}
V.~A. Novikov, Mikhail~A. Shifman, A.~I. Vainshtein, and Valentin~I. Zakharov.
\newblock {Exact Gell-Mann-Low Function of Supersymmetric Yang-Mills Theories
  from Instanton Calculus}.
\newblock {\em Nucl. Phys.}, B229:381, 1983.

\bibitem{jones}
D.~R.~T. Jones.
\newblock {MORE ON THE AXIAL ANOMALY IN SUPERSYMMETRIC YANG-MILLS THEORY}.
\newblock {\em Phys. Lett.}, B123:45, 1983.

\bibitem{Seiberg:1994bp}
Nathan Seiberg.
\newblock {The Power of holomorphy: Exact results in 4-D SUSY field theories}.
\newblock 1994.

\bibitem{svabelian}
A.~I. Vainshtein, Valentin~I. Zakharov, and Mikhail~A. Shifman.
\newblock {Gell-Mann Low Function in Supersymmetric Electrodynamics}.
\newblock {\em JETP Lett.}, 42:224--227, 1985.

\bibitem{dineshirman}
Michael Dine and Yuri Shirman.
\newblock {Some explorations in holomorphy}.
\newblock {\em Phys. Rev.}, D50:5389--5397, 1994.

\bibitem{higherderivative1}
A.~B. Pimenov, E.~S. Shevtsova, A.~A. Soloshenko, and K.~V. Stepanyantz.
\newblock {Higher derivative regularization and quantum corrections in N=1
  supersymmetric theories}.
\newblock 2007.

\bibitem{higherderivative2}
A.~B. Pimenov, E.~S. Shevtsova, and K.~V. Stepanyantz.
\newblock {Calculation of two-loop beta-function for general N=1 supersymmetric
  Yang--Mills theory with the higher covariant derivative regularization}.
\newblock {\em Phys. Lett.}, B686:293--297, 2010.

\bibitem{littlemiracles}
Mikhail~A. Shifman.
\newblock {Little Miracles of Supersymmetric Evolution of Gauge Couplings}.
\newblock {\em Int. J. Mod. Phys.}, A11:5761--5784, 1996.

\bibitem{shifmanweakcoupling}
Mikhail~A. Shifman and A.I. Vainshtein.
\newblock {On Gluino Condensation in Supersymmetric Gauge Theories. SU(N) and
  O(N) Groups}.
\newblock {\em Nucl.Phys.}, B296:445, 1988.

\bibitem{fuchs}
J.~Fuchs and M.~G. Schmidt.
\newblock {Instanton induced Green functions in the superfield formalism}.
\newblock {\em Z. Phys.}, C30:161, 1986.

\bibitem{fp}
D.~Finnell and P.~Pouliot.
\newblock {Instanton calculations versus exact results in four- dimensional
  SUSY gauge theories}.
\newblock {\em Nucl. Phys.}, B453:225--239, 1995.

\bibitem{hollowood}
Timothy~J. Hollowood, Valentin~V. Khoze, Weon-Jong Lee, and Michael~P. Mattis.
\newblock {Breakdown of cluster decomposition in instanton calculations of the
  gluino condensate}.
\newblock {\em Nucl. Phys.}, B570:241--266, 2000.

\bibitem{Dorey:2002ik}
Nick Dorey, Timothy~J. Hollowood, Valentin~V. Khoze, and Michael~P. Mattis.
\newblock {The calculus of many instantons}.
\newblock {\em Phys. Rept.}, 371:231--459, 2002.

\bibitem{vainshtein}
Adam Ritz and Arkady~I. Vainshtein.
\newblock {Instantons at strong coupling, averaging over vacua, and the gluino
  condensate}.
\newblock {\em Nucl.Phys.}, B566:311--328, 2000.

\bibitem{ArkaniHamed:1997ut}
Nima Arkani-Hamed and Hitoshi Murayama.
\newblock {Renormalization group invariance of exact results in supersymmetric
  gauge theories}.
\newblock {\em Phys. Rev.}, D57:6638--6648, 1998.

\bibitem{ArkaniHamed:1997mj}
Nima Arkani-Hamed and Hitoshi Murayama.
\newblock {Holomorphy, rescaling anomalies and exact beta functions in
  supersymmetric gauge theories}.
\newblock {\em JHEP}, 06:030, 2000.

\bibitem{johansen}
A.~A. Johansen.
\newblock {ULTRAVIOLET REGULARIZATION OF N=1 SUSY GAUGE THEORIES BY EXTENDED
  SUSYS}.
\newblock {\em Yad. Fiz.}, 45:263--274, 1987.

\bibitem{parkes}
A.~Parkes and Peter~C. West.
\newblock {FINITENESS AND EXPLICIT SUPERSYMMETRY BREAKING OF THE N=4
  SUPERSYMMETRIC YANG-MILLS THEORY}.
\newblock {\em Nucl.Phys.}, B222:269, 1983.

\bibitem{Kovner:1997im}
A.~Kovner and Mikhail~A. Shifman.
\newblock {Chirally symmetric phase of supersymmetric gluodynamics}.
\newblock {\em Phys. Rev.}, D56:2396--2402, 1997.

\bibitem{davies}
N.~Michael Davies, Timothy~J. Hollowood, Valentin~V. Khoze, and Michael~P.
  Mattis.
\newblock {Gluino condensate and magnetic monopoles in supersymmetric
  gluodynamics}.
\newblock {\em Nucl. Phys.}, B559:123--142, 1999.

\bibitem{Dine:2010jg}
Michael Dine, Guido Festuccia, Lawrence Pack, and Weitao Wu.
\newblock {Reliable Semiclassical Computations in QCD}.
\newblock {\em Phys. Rev.}, D82:065015, 2010.

\bibitem{amati}
D.~Amati, G.~C. Rossi, and G.~Veneziano.
\newblock {Instanton Effects in Supersymmetric Gauge Theories}.
\newblock {\em Nucl. Phys.}, B249:1, 1985.

\bibitem{Diakonov:2002qw}
Dmitri Diakonov and Victor Petrov.
\newblock {Gluino condensate and long range fields}.
\newblock {\em Phys.Rev.}, D67:105007, 2003.

\bibitem{betadependence}
N.~Dorey, D.~Tong, and S.~Vandoren.
\newblock {Instanton effects in three-dimensional supersymmetric gauge theories
  with matter}.
\newblock {\em JHEP}, 9804:005, 1998.

\bibitem{leeinstanton}
Ki-Myeong Lee and Chang-hai Lu.
\newblock {SU(2) calorons and magnetic monopoles}.
\newblock {\em Phys. Rev.}, D58:025011, 1998.

\bibitem{Kraan:1998kp}
Thomas~C. Kraan and Pierre van Baal.
\newblock {Exact T-duality between calorons and Taub - NUT spaces}.
\newblock {\em Phys. Lett.}, B428:268--276, 1998.

\bibitem{Weinberg:2006rq}
Erick~J. Weinberg and Piljin Yi.
\newblock {Magnetic Monopole Dynamics, Supersymmetry, and Duality}.
\newblock {\em Phys.Rept.}, 438:65--236, 2007.

\bibitem{Wess:1992cp}
J.~Wess and J.~Bagger.
\newblock {Supersymmetry and supergravity}.
\newblock Princeton, USA: Univ. Pr. (1992) 259 p.

\bibitem{Creamer:1977vf}
Dennis~B. Creamer.
\newblock {Scalar Propagators in a Pseudoparticle Field}.
\newblock {\em Phys. Rev.}, D16:3496, 1977.

\end{thebibliography}
\bibliographystyle{unsrt}
\end{document}